 \documentclass[preprint,3p,times,twocolumn]{elsarticle}

\usepackage{amssymb}
\usepackage{amsmath}

\usepackage{xcolor}

\usepackage{natbib}
\newcommand{\mev}{MeV}

\newcommand{\eg}{{\it e.g.}}
\newcommand{\ie}{{\it i.e.}}

\newcommand{\rhoem}{\varrho_{\rm EM}}
\newcommand{\Piem}{\Pi_{\rm EM}}

\newcommand{\lsim}{\lesssim}
\newcommand{\gsim}{\gtrsim}
\newcommand{\beq}{\begin{equation}}
\newcommand{\eeq}{\end{equation}}
\newcommand{\bea}{\begin{eqnarray}}
\newcommand{\eea}{\end{eqnarray}}
\newcommand{\bef}{\begin{figure}}
\newcommand{\eef}{\end{figure}}
\newcommand{\bce}{\begin{center}}
\newcommand{\ece}{\end{center}}
\newcommand{\fB}{f_{\mathrm{B}}}

\journal{Physics Letters B}

\begin{document}
\begin{frontmatter}



\title{Polarization of Thermal Dilepton Radiation}


\author[tud]{Florian Seck\corref{fs}}
\cortext[fs]{Corresponding author.}
\ead{f.seck@gsi.de}
\affiliation[tud]{organization={Technische Universität Darmstadt}, postalcode={64289}, city={Darmstadt}, country={Germany}}
\author[gsi,tud]{Bengt Friman}
\affiliation[gsi]{organization={GSI Helmholtzzentrum für Schwerionenforschung GmbH}, postalcode={64291}, city={Darmstadt}, country={Germany}}
\author[tud,gsi]{Tetyana Galatyuk}
\author[gu,helm]{Hendrik van Hees}
\affiliation[gu]{organization={Institut für Theoretische Physik, Goethe-Universität Frankfurt am Main}, addressline={Max-von-Laue-Strasse 1}, postalcode={60438}, city={Frankfurt am Main}, country={Germany}}
\affiliation[helm]{organization={Helmholtz Research Academy Hesse for FAIR}, addressline={Campus Frankfurt}, postalcode={60438}, city={Frankfurt}, country={Germany}}
\author[tamu]{Ralf Rapp}
\affiliation[tamu]{organization={Texas A$\&$M University}, city={College Station}, postalcode={77843-3366}, state={TX}, country={USA}}
\author[illi]{Enrico Speranza\fnref{es}}
\fntext[es]{Present address: Theoretical Physics Department, CERN, 1211 Geneva 23, Switzerland}
\affiliation[illi]{organization={Illinois Center for Advanced Studies of the Universe \& Department of Physics, University of Illinois Urbana-Champaign}, city={Urbana}, postalcode={61801}, state={IL}, country={USA}}
\author[tud]{Jochen Wambach}
%

\begin{abstract}
The invariant mass spectra of dileptons radiated from the fireballs formed in high-energy heavy-ion collisions have been successfully used to investigate the properties of hot and dense QCD matter. Using a realistic model for the in-medium electromagnetic spectral function, we predict polarization observables and compare them to experiment. This allows, for the first time, independent tests of the longitudinal and transverse components of the virtual photon's selfenergy.
While the low- and high-mass regions exhibit the expected limits of transverse and unpolarized photons, respectively, baryon-driven medium effects in the $\rho$-meson mass region create a marked longitudinal polarization that transits into a largely unpolarized emission from the quark-gluon plasma, thus providing a sensitive test of microscopic emission processes in QCD matter. Applications to available data from the HADES and NA60 experiments at SIS and SPS energies, respectively, are consistent with our predictions and set the stage for quantitative polarization studies at FAIR and collider energies.
\end{abstract}

\begin{keyword}
heavy-ion collisions \sep QCD matter \sep dilepton radiation \sep polarization observables

\end{keyword}

\end{frontmatter}


\section{Introduction}
%
Measurements of electromagnetic (EM) radiation from ultrarelativistic heavy-ion collisions (URHICs) have provided unprecedented insights into the properties of Quantum Chromodynamics (QCD) matter formed in these reactions. Over the last decade, a rather consistent picture has emerged in interpreting the observed dilepton spectra. At low invariant masses, commonly referring to $M\lsim1$\,GeV, thermal radiation mostly emanates from the hadronic medium of the fireball evolution, with a strongly broadened $\rho$-meson peak indicating an ultimate melting and transition into a continuum of partonic degrees of freedom~\cite{CERES:2005uih,NA60:2006ymb}. Similar findings have also been reported at the higher energies of the Relativistic Heavy-Ion Collider (RHIC)~\cite{STAR:2013pwb} and the lower energies at the Schwerionensynchrotron (SIS18)~\cite{HADES:2011nqx}. In the intermediate-mass region, 1\,GeV~$\lsim~M\lsim$~3\,GeV, the radiation is strongly favored toward early phases, which, at least for collision energies of $\sqrt{s}\gsim$\,10\,GeV, has been associated with partonic radiation sources~\cite{Rapp:2014hha}, with temperatures well above the pseudocritical transition temperature obtained from lattice QCD (lQCD), $T_{\rm pc}$\,$\simeq$\,155-160\,MeV~\cite{HotQCD:2018pds,Borsanyi:2020fev}.

Successful model descriptions of dilepton data largely rely on hadronic many-body theory, where the predicted melting of the $\rho$-meson rather seamlessly transits into a structureless quark-antiquark ($q\bar{q}$) continuum~\cite{Rapp:1999us,Peters:1997va,Hayano:2008vn}, albeit with substantial enhancements over the free $q\bar q$ rate toward low masses. However, the precise micro-physics underlying the strongly coupled QCD liquid in the transition regime remains a matter of debate. Therefore, further tests of the existing model calculations would be very valuable. In this letter we demonstrate this in a first quantitative application to spin-polarization observables of low-mass dileptons in heavy-ion experiments. Polarization observables in URHICs have recently garnered attention in a broader context, \eg, in the detection of the ``most vortical fluid'' or the production of unprecedented magnetic fields~\cite{STAR:2017ckg,ALICE:2022dyy,STAR:2022fan}. Polarization of dileptons, as a penetrating probe, is expected to play a key role also in these efforts, and it is therefore important to have good control over its thermal emission characteristics.

%
\section{Methodology of dilepton polarization}

\begin{figure*}[tbh]
    \centering
    \includegraphics[width=0.95\textwidth]{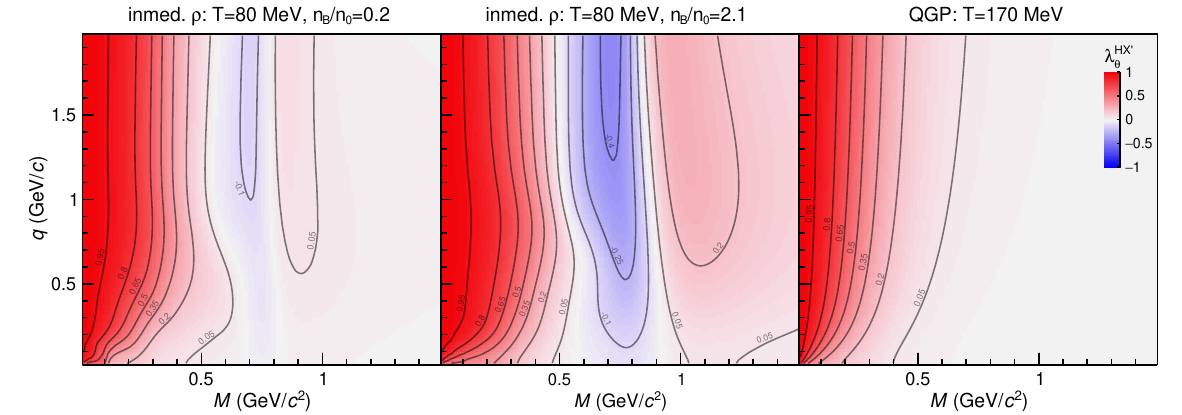}
    \caption{Anisotropy coefficient, $\lambda_{\theta}^{\rm HX'}$ from Eq.~(\ref{lambda_rest}), for EM spectral functions in static hadronic matter at $T$\,=\,80\,\mev~with baryon densities $n_{B}$\,=\,0.2(2.1)$\,n_{0}$ in the left (middle) panel ($n_{0}$\,=\,0.16\,fm$^{-3}$), and in a static QGP at $T$\,=\,170\,\mev~(right panel). }
    \label{fig:lambda}
\end{figure*}

The key quantity in our study is the EM emissivity of equilibrium QCD matter at temperature $T$ and baryo-chemical potential $\mu_{\mathrm{B}}$ which is determined by the thermal expectation value of the EM current correlator, $\Piem^{\mu\nu}$. It can also be interpreted as the in-medium selfenergy of a photon with four-momentum $q^\mu=(q_0,\vec q)$. The pertinent spectral function, $\rhoem^{\mu\nu}$=$-2\, {\rm Im}{\Piem^{\mu\nu}}$, figures in the dilepton emission rate as
\begin{equation}
\frac{dN_{ll}}{d^4x\,d^4q}=\frac{\alpha^2 L(M)}{6\pi^3M^2}
\fB(q_0;T)g_{\mu\nu}\rhoem^{\mu\nu}(M,|\vec{q}|;T,\mu_{\mathrm{B}}) \ ,
\label{rate}
\end{equation}
where $M$=$\sqrt{q_0^2- \vec q^2}$ denotes the dilepton invariant mass, $\fB(q_0;T)=1/(e^{q_0/T}-1)$ the thermal Bose function, 
$\alpha\simeq 1/137$ the fine-structure constant and $L(M)$ a lepton phase-space factor ($L(M)$=1 for invariant mass $M \gg m_l$).
Using the 4-dim projectors for a spin-1 particle, $P^{\mu\nu}_{L,T}$, one can decompose the spectral function into its longitudinal and transverse components as~\cite{McLerran:1984ay,Landsman:1986uw}
\begin{equation}
\rhoem^{\mu\nu} = \varrho_L P_L^{\mu\nu} + \varrho_T P_T^{\mu\nu} \ ,
\end{equation}
rendering $g_{\mu\nu}\rhoem^{\mu\nu} =\varrho_L+2\varrho_T$. At vanishing three-momentum relative to the heat bath, one has $\varrho_T=\varrho_L$ for all $M$. At finite $|\vec{q}|$, this no longer holds as spherical symmetry is broken. For real photons ($M$=0),
$\varrho_L$ vanishes, and the emission rate is purely transverse.   

Angular dependencies in the dilepton production rate can be unraveled by resolving the lepton angle, ${\Omega_l=(\phi_l,\theta_l)}$,  in the photon rest frame~\cite{Bratkovskaya:1995kh,Baym:2017qxy,Speranza:2018osi}. Using
\begin{equation}
\frac{d N_{ll}}{d^4x\, d^4q\, d\Omega_l}= \frac{\alpha^2}{32 \pi^4}\frac{1}{M^4}\sqrt{1-\frac{4 m_l^2}{M^2}}\rhoem^{\mu\nu}L_{\mu\nu} \fB(q_0;T)\ ,
\end{equation}
with the lepton tensor 
\begin{equation}
L^{\mu\nu}=2(q^2 g^{\mu\nu} - q^\mu q^\nu + \Delta l^\mu \Delta l^\nu) \ ,
\end{equation}
($\Delta l^\mu = l^{+\mu}-l^{-\mu}$, $l^{\pm}$:lepton four-momenta), the angular distribution can be expressed as
\begin{align}
\label{angdist}
\frac{d N_{ll}}{d^4 x \, d^4q \,d\Omega_l}&\propto
\frac{1}{3+\lambda_\theta}
\Big(1+\lambda_\theta\cos^2\theta_l\\
&+\lambda_\phi \sin^2\theta_l\cos2\phi_l+\lambda_{\theta\phi}\sin2\theta_l\cos\phi_l \nonumber\\
&+\lambda^{\bot}_\phi \sin^2\theta_l\sin2\phi_l+\lambda^{\bot}_{\theta\phi}\sin2\theta_l\sin\phi_l \Big)\nonumber \ ,
\end{align}
where the $\lambda$'s are the anisotropy coefficients.

Well-known examples in this context are the Drell-Yan process~\cite{Drell:1970wh} (leading to a transverse polarization with respect to the relative momentum of the incoming quark and antiquark) and the high-$p_T$ $J/\psi$ polarization puzzle~\cite{Faccioli:2010kd,Feng:2018ukp} (where gluon fragmentation suggests transverse polarization, at variance with experiment).
Even in an isotropic thermal medium, finite anisotropies in the angular distribution of the produced leptons occur, \eg, for basic hadronic and partonic sources ($\pi\pi$ 
and $q\bar q$ annihilation, respectively) at the few percent level~\cite{Speranza:2018osi}. However, the effects are expected to be much larger at smaller $M$, where more intricate production processes, such as resonance Dalitz decays and Bremsstrahlung, contribute and smoothly approach fully transverse polarization at the photon point.
Moreover, the anisotropy of lepton pairs may help to disentangle the sources in the $M$=1-1.5\,GeV region, where ``chiral mixing" between the $\rho$ and $a_1$ channels via $\pi a_1$ annihilation \cite{Hohler:2013eba, Jung:2016yxl} competes with $q \bar q$ annihilation.

\section{Dilepton angular distribution with realistic spectral functions}

In the present study, we employ in-medium spectral functions that provide a satisfactory description of available dilepton data from ultrarelativistic heavy-ion collisions. They account for hadronic emission with in-medium vector-meson (mostly $\rho$-meson) spectral functions calculated in hadronic many-body theory based on effective Lagrangians~\cite{Rapp:1997fs,Rapp:1999us}, supplemented with continuum-like multi-meson annihilation channels relevant at masses above $\sim$1\,GeV~\cite{vanHees:2007th}, and emission from a quark-gluon plasma (QGP) based on $q\bar q$ annihilation with a low-energy transport peak constrained by lQCD data~\cite{Rapp:2013nxa}.
The transition from hadronic to QGP radiation is carried out at a temperature of 170\,MeV, where the two rates are close to each other. 


Let us start by inspecting the pertinent anisotropy coefficients in a static thermal medium. It is characterized by the four-velocity ${u^{\mu}}'=(1,0,0,0)$, and the rotational symmetry is broken only by the three-momentum $\vec q$ of the virtual photon. The helicity frame $\textrm{HX'}$ is defined in the rest frame of the photon, choosing the polar axis $z'$ along $\vec q$.
  
In the limit of small lepton masses, $m_l\ll M$ (which we will employ for the remainder of the manuscript), the only non-vanishing anisotropy coefficient $\lambda_{\theta}$ is given by
\begin{equation}
    \lambda_{\theta}^{\rm HX'} (M,|\vec{q}|)
    \underset{m_l \ll M}{\cong} 
    \frac{ \varrho_{\rm T} - \varrho_{\rm L} } { \varrho_{\rm T} + \varrho_{\rm L} } \ , 
    \label{lambda_rest}
\end{equation}
which highlights its dependence on the {\it difference} between the polarization components of the EM spectral function. The results for hadronic matter and the QGP are displayed in Fig.~\ref{fig:lambda}.
All other anisotropy coefficients vanish as they involve asymmetries with respect to the $\phi_l$ angle of the leptons.
The hadronic spectral function exhibits a strong dependence of $\lambda_{\theta}$ on the baryon density and, in particular, a non-monotonic one on invariant mass, $M$. By contrast, the polarization of QGP radiation is rather small, except when approaching the photon point, $M\to 0$, where $\lambda_\theta$ becomes unity. Thus, in this case the polarization exhibits a smooth transition from transversely polarized photons for $M\lesssim 0.5$ GeV with $M/q \ll 1$ to unpolarized emission at high masses. Medium effects in hadronic emission at high $\mu_{\mathrm{B}}$ generate remarkable longitudinal polarization around the $\rho$ mass. It is intriguing to note that the source of this longitudinal polarization is (at least in part) the coupling of the photon to negative-parity baryon resonances, such as $N(1520)$, $N(1535)$, and $\Delta(1700)$, which are chiral partners of the nucleon and $\Delta(1232)$~\cite{Rapp:2002tw}. The implications for a signal of chiral symmetry restoration through a ``parity doubling'' of baryons~\cite{Tripolt:2021jtp} require further scrutiny.


\begin{figure*}[tbh]
    \centering
    \includegraphics[width=0.65\textwidth]{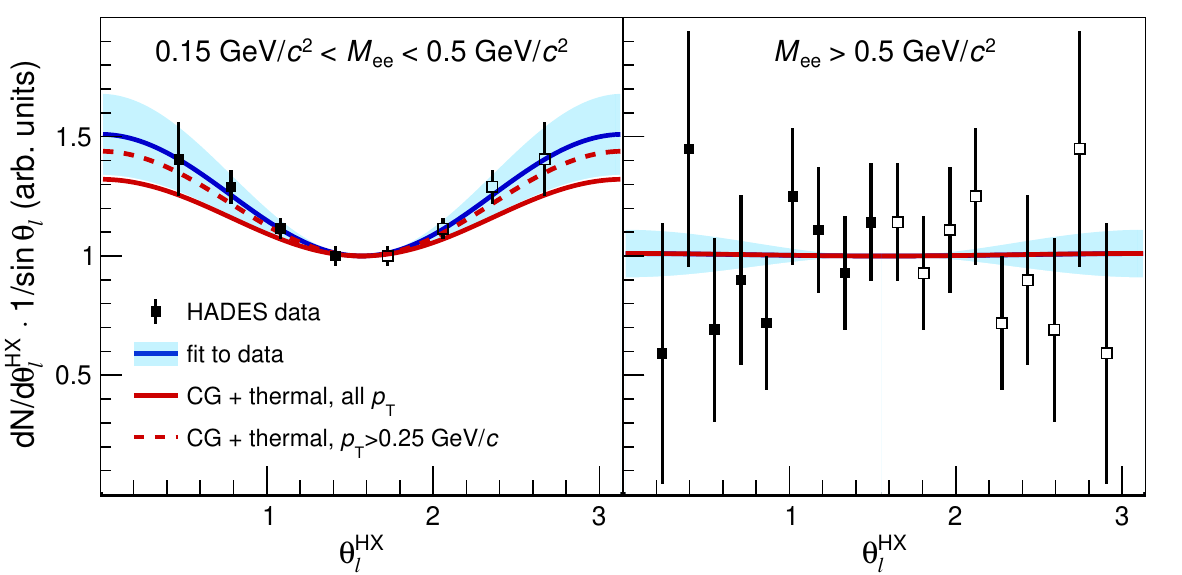}
    \caption{Dielectron angular distributions (red lines) calculated from coarse-grained transport model simulations combined with thermal rates, integrated over two invariant mass bins, compared to HADES data in Ar(1.76\,AGeV)+KCl collisions (full squares; open data points are reflected)~\cite{HADES:2011nqx}. Fits by the HADES collaboration (blue lines with bands) yield anisotropy parameters of $\lambda_{\theta}$=$0.51 \pm 0.17$ (left panel) and $0.01\pm 0.1$ (right panel) for the lower and higher mass window, respectively, compared to 0.32 and 0.01 from the present calculations. To simulate the effects of a finite HADES acceptance, we also show our results with a cut on transverse pair momentum in the lab frame of $p_T>0.25$\,GeV (red dashed line).}
    \label{fig:hades}
\end{figure*}

To evaluate dilepton radiation in a heavy-ion collision, one generally divides the expanding fireball into small cells of locally thermalized matter with velocity, $u^\mu=\gamma\,(1,\vec{\beta})$, in the center-of-mass (CM) frame of the collision (as is routinely done in hydrodynamic simulations).
To compute the polarization of a lepton pair within each cell, the local helicity frame HX', which, in analogy to the static case, is reached by a Lorentz-boost from the rest frame of the fluid cell to the rest frame of the virtual photon. 
The task is now to transform the local polarization in the HX' frame (where only $\lambda_{\theta}^{\rm HX'}$ is non-vanishing) into a global frame accessible to experiments (where all anisotropy coefficients may acquire non-zero values). Common choices are the helicity (HX) and Collins-Soper (CS) frames \cite{Collins:1977iv}.  In the latter, the polar axis is given by the bisector of the angle formed by the beam momenta in the rest frame of the virtual photon. Since these reference frames are all defined in the rest frame of the virtual photon, they differ only by the orientation of the coordinate system. Consequently, the transformations between them involve only rotations in three dimensions.

We start from a virtual photon with the (observed) four-momentum $p^{\mu}=(p_0,0,0,p)$ in the CM system of the collision, defining the $z$-axis along its three-momentum, $\vec{p}$. 
In the helicity frame (HX), this is chosen as the polar axis, while the
corresponding $y$-axis is defined along the normal vector of the plane spanned by the beam momenta, thus completely specifying the ($xyz$) system. On the other hand, the $z'$ axis in the ${\rm HX'}$ frame is chosen along the photon momentum in the thermal rest frame of the emitting fluid cell, which is moving with the flow velocity, $u^\mu$, in the CM system. The photon four-momentum in that frame, $q^\mu$, is obtained by a Lorentz-boost of $p^\mu$ using $u^\mu$. This determines the only non-zero coefficient in the ${\rm HX'}$ system, $\lambda_{\theta}^{\rm HX'}$, as outlined above. With these definitions, one can then transform the angular distribution into the HX system by three successive Euler rotations~\cite{Faccioli:2022peq}: \\
    (i) around the $z'$-axis by an angle $\psi$ to bring the $y$-axis perpendicular to the $z$-axis;\\
    (ii) around the thus obtained $y''$ axis by an angle $\zeta$ to align the $z'$-axis with the $z$-axis; and\\ 
    (iii) around the $z$-axis by an angle $\omega$ to align the $x'$- and $y'$-axes along the $x$- and $y$-axes, respectively. \\
In this way, all five coefficients in Eq.~(\ref{angdist}) can be determined from $\lambda_{\theta}^{\rm HX'}$ and the three rotation angles described above.
A similar procedure, to be detailed elsewhere, can be carried out when measuring the polarization in the Collins-Soper frame.


To compare with experiment,  the contributions to the anisotropy coefficients in a given frame (\eg, HX) from all fluid cells need to be accounted for. In practice, this is achieved via a  yield-weighted mean~\cite{Faccioli:2022peq},
\begin{equation}
    \lambda_{\theta}^{\rm HX} (M,p_T,y,\phi) =
    \frac{ \displaystyle\sum_{\rm all \; cells} \mathcal{N}_{\rm cell} \; \lambda_{\theta, \rm cell}^{\rm HX} }{ \mathcal{N}_{\rm fireball} } \ ,
\end{equation}
where $\mathcal{N}_{\rm fireball} = \displaystyle\sum_{\rm all \; cells} \mathcal{N}_{\rm cell}$ with a cell weight 
\begin{equation}
    \mathcal{N}_{\rm cell} =
    \frac{ dN_{ll}( T_{\rm cell}, \mu_{B,{\rm cell}}) }{ d^4x \, dM \, dp_T \, dy \, d\phi } \times \frac{V_{\rm cell} \, t_{\rm cell}} { 3+\lambda_{\theta, \rm cell}^{\rm HX} } \ .
\end{equation}
Here, $V_{\rm cell}$ denotes the cell's three-volume and $t_{\rm cell}$ the time discretization interval. Note that 
\begin{equation}
    \frac{ dN_{ll}( T_{\rm cell}, \mu_{B,{\rm cell}}) }{d^4x \, dM \, dp_T \, dy \, d\phi } = M \, p_T \, \frac{ dN_{ll}( T_{\rm cell}, \mu_{B,{\rm cell}}) }{d^4x \, d^4q } \ ,
\end{equation}
where $dN_{ll}( T_{\rm cell}, \mu_{B,{\rm cell}})/(d^4x\, d^4q)$  is a Lorentz-invariant. In a given experiment, limited statistics may be compensated for by increasing the bin size, \ie, by integrating over kinematic variables. For the anisotropy coefficient $\lambda_{\theta}^{\rm HX} (M)$, as a function of invariant mass, one finds
\begin{equation}
    \begin{split}
        \lambda_{\theta}^{\rm HX} (M) &= 
        \frac{ \int \mathcal{N}_{\rm fireball} \; \lambda_{\theta}^{\rm HX} (M,p_T,y,\phi) \; dp_T \; dy \; d\phi }{ \int \mathcal{N}_{\rm fireball} \; dp_T \; dy \; d\phi } \\ &= 
        \frac{ \sum \mathcal{N}_{\rm fireball} \; \lambda_{\theta}^{\rm HX} (M,p_T,y,\phi) \; \Delta p_T \; \Delta y \; \Delta \phi }{ \sum \mathcal{N}_{\rm fireball} \; \Delta p_T \; \Delta y \; \Delta \phi } \ ,       
    \end{split}
\end{equation}
where $\Delta p_T$, $\Delta y$, and $\Delta \phi$ are the bin widths for transverse momentum, rapidity, and azimuthal angle, respectively.

\begin{figure*}[thb]
    \centering
    \vspace{0.3cm}
    \includegraphics[width=0.65\textwidth]{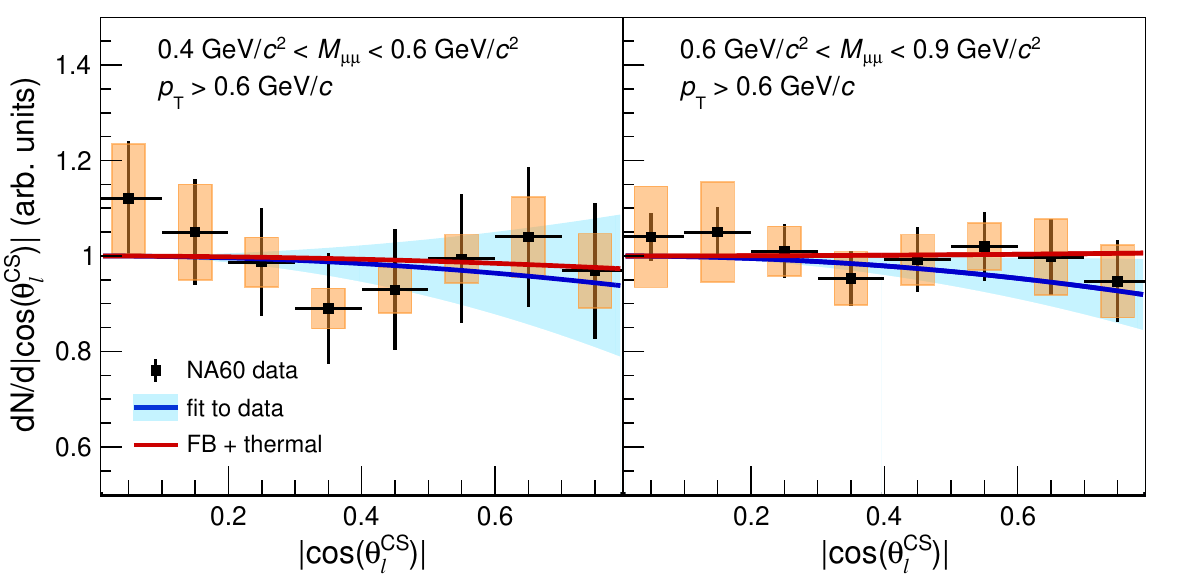}\\
    \vspace{0.3cm}
    \includegraphics[width=0.65\textwidth]{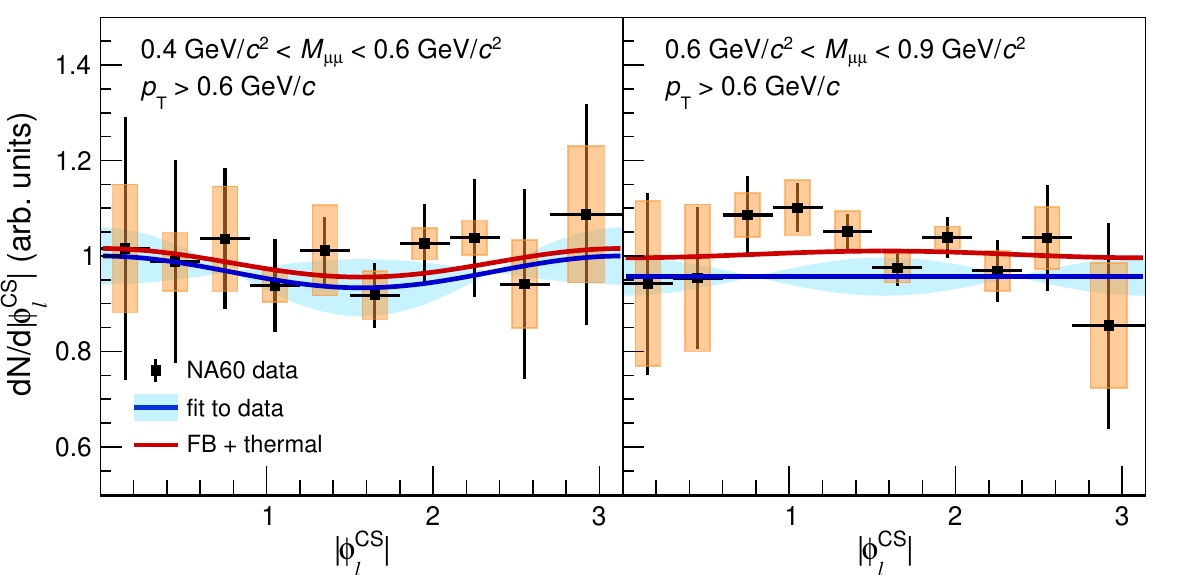}
    \caption{Dimuon angular distributions (red lines) calculated from a fireball (FB) model combined with thermal rates, integrated over two dimuon mass windows, in the Collins-Soper frame, compared to NA60 data in In(158\,AGeV)+In collisions~\cite{NA60:2008iqj}. The blue lines represent fits by NA60 with extracted anisotropy parameters of $\lambda_{\theta}$= -$0.10\pm0.24$ (upper left) and -$0.13\pm0.12$ (upper right), compared to the calculated values of -0.04 and 0.01, respectively, and likewise $\lambda_{\phi}$=$0.05\pm0.09$ (lower left) and $0.00 \pm 0.06$ (lower right), compared to 0.04 and -0.01 from our calculation.}
    \label{fig:InIn_data_comparison}
\end{figure*}

\section{Results and comparison to experimental data}

We first compare our results to HADES data~\cite{HADES:2011nqx} measured 
in Ar(1.76\,AGeV)+KCl collisions, cf.~Fig~\ref{fig:hades}. For this collision system, the discretization into space-time cells follows our previous work~\cite{Galatyuk:2015pkq} employing a coarse-graining of the UrQMD transport model without the formation of a QGP phase. For both invariant mass bins shown in the two panels, the theoretical predictions (red lines) are in quite good agreement with the data and not far from the functional best fit (blue lines), in particular, when a cut in transverse pair momentum approximately accounts for experimental acceptance (other theoretical uncertainties are estimated to be much smaller).
We conclude that the main characteristics of the anisotropy coefficient $\lambda_{\theta}$ in a static thermal medium survive the transformations induced by the fireball expansion. This is not surprising since, after all, the collective flow developed in Ar(1.76\,AGeV)+KCl collisions is rather small. Consequently, the data directly reflect the polarization properties of the hadronic EM spectral function shown in the left and middle panels of Fig.~\ref{fig:lambda}, which exhibits a transition from transverse polarization for masses below 0.5\,GeV (red region) to a regime where the (average) polarization is small.

Next, we turn to the NA60 dimuon data~\cite{NA60:2008iqj} from In(158\,AGeV)+In
collisions. For this system, we utilize the isentropically expanding fireball model whose particle content reproduces the experimentally observed light-hadron data and whose time evolution was constrained by observed $p_T$ spectra and hydrodynamic expansion time scales~\cite{vanHees:2007th,Rapp:2014hha}. 
This model includes a QGP phase that reaches maximal initial temperatures close to $T_0=240$\,MeV. Also here, the theoretical predictions (in this case in the Collins-Soper frame) describe the measured angular distributions in $\theta_l$ and $\phi_l$ quite well, cf.~Fig.~\ref{fig:InIn_data_comparison}.
Note, however, that according to our calculation, the near absence of a net polarization (\ie, the rather flat angular distributions) is not {\it a priori} related to thermal isotropy arguments but stems from the properties of the equilibrium EM spectral function, \ie, the rather small net polarization in the mass region around $\sim$0.5\,GeV (see the right-most panel in Fig.~\ref{fig:lambda}). 

\section{Summary and conclusions}

In summary, we have extracted polarization properties of virtual photons from theoretical electromagnetic spectral functions for QCD matter that provide a fair agreement with existing dilepton data for mass and momentum spectra. 
While such spectra rely on the sum, $2\varrho_T+\varrho_L$, of the transverse and longitudinal components of the spectral function, polarization observables are sensitive to their difference, $\varrho_T-\varrho_L$. We find that the virtual photons emitted by dense hadronic matter exhibit a structure where a transverse polarization at low masses is converted into a longitudinal one in the $\rho$-meson mass region. This effect can be traced back largely to the excitation of baryonic resonances.  At high temperatures and smaller baryon densities, these structures become less pronounced and more closely resemble emission characteristics for a QGP.
We have applied these model predictions to experimental observables by carrying out the transformations from the local thermal frames into angular variables observable in the lab frame. Our predictions agree fairly well with both HADES and NA60 data, supporting the microscopic description underlying our model.
In the future, we expect polarization observables to play an increasingly important role in exploring the mechanisms underlying dilepton emission spectra in heavy-ion collisions. Multi-differential measurements of the virtual photon polarization will become available from HADES, STAR, and ALICE, as well as from the future high-rate experiments CBM, NA60+, and \mbox{ALICE3}. 

\section*{Acknowledgments}
This work was supported by the Deutsche Forschungsgemeinschaft (DFG, German Research Foundation) through
project number 315477589 – TRR 211 ``Strong-interaction matter under extreme conditions'', by the U.S. National Science Foundation (NSF) under grant no. PHY-2209335 and the ExtreMe Matter Institute EMMI at the GSI Helmholtzzentrum für Schwerionenforschung, Darmstadt, Germany (RR).

\bibliography{bibnew}
\end{document}